\documentclass[%
 reprint,
 amsmath,amssymb,
 aps,
 pre
]{revtex4-1}
\renewcommand\appendix{\par 
   \setcounter{section}{0}%
   \setcounter{subsection}{0}%
   \setcounter{figure}{0}%
   \renewcommand\thesection{\Alph{section}}%
   \renewcommand\thefigure{\Alph{section}.\arabic{figure}}} 

\usepackage{graphicx}
\usepackage{dcolumn}
\usepackage{bm}

\begin{document}

\title{Langevin Dynamics simulations of a 2-dimensional colloidal crystal under confinement and shear}
\author{D. Wilms$^{1,2}$, P. Virnau$^1$, S. Sengupta$^{3,4}$ and K. Binder$^1$}
\affiliation{$^1$Institut f\"ur Physik, Johannes Gutenberg-Universit\"at Mainz, Staudinger Weg 7, D-55099 Mainz, Germany}
\affiliation{$^2$Graduate School Materials Science in Mainz, Staudinger Weg 9, D-55128 Mainz, Germany}
\affiliation{$^3$Indian Association for the Cultivation of Science, Centre for Advanced Materials, Jadavpur, Kolkata 700032, India}
\affiliation{$^4$Advanced Materials Research Unit, S. N. Bose National Centre for Basic Science, Block JD, Sector III, Salt Lake, Kolkata 700098, India}
\begin{abstract}

Langevin Dynamics simulations are used to study the effect of shear on a two-dimensional colloidal crystal confined by structured parallel walls. When walls are sheared very slowly, only two or three crystalline layers next to the walls move along with them, while the inner layers of the crystal are only slightly tilted. At higher shear velocities, this inner part of the crystal breaks into several pieces with different orientations. The velocity profile across the slit is reminiscent of shear-banding in flowing soft materials, where liquid and solid regions coexist; the difference, however, is that in the latter case the solid regions are glassy while here they are crystalline. At even higher shear velocities, the effect of the shearing becomes smaller again. Also the effective temperature near the walls (deduced from the velocity distributions of the particles) decreases again when the wall velocity gets very large. When the walls are placed closer together, thereby introducing a misfit, a structure containing a soliton staircase arises in simulations without shear. Introducing shear increases the disorder in these systems until no solitons are visible any more. Instead, similar structures like in the case without misfit result. At high shear rates, configurations where the incommensurability of the crystalline structure is compensated by the creation of holes become relevant.

\end{abstract}
\maketitle

\section{Motivation}

Colloidal particles are a very useful model system for the study of fundamental properties of condensed matter, for example for the study of disorder and statistical fluctuations \cite{2,3,4,5}. These particles are convenient, because their interaction is tunable in many ways. For instance, by adding variable concentrations of non-adsorbing polymers to a suspension of colloidal particles, their hard-core repulsion is amended by a depletion attraction whose strength is controlled by the polymer concentration. Another example are charged colloids, where adding salt can modify the strength and range of the effective Yukawa potential between the particles. Finally, if one uses colloidal particles that contain a superparamagnetic core, a magnetic field can be used to control the strength of the $r^{-3}$-dipole-dipole-potential acting between them \cite{2,3,4,5,1,6,12,13}. Their size also makes them useful for experiments as it typically lies in a range which can be observed directly by confocal miscroscopy. Additionally, the size and shape of these particles can be varied. It is also feasible to trap them in layers \cite{1,2,3,4,5,6,7,8,9,10,11}, e.g. at air-water interfaces, in order to get a two-dimensional system. Such systems can be confined in a mechanical way or by laser fields \cite{7,8,14,15}. The effect of shear on the homogeneous and heterogeneous nucleation of colloidal suspensions \cite{71,72,73,74}, on phase separation \cite{75} and on colloidal glasses \cite{76} has also been adressed.

One of the important questions in this context is how the structure formation in colloidal systems can be influenced by the various tunable parameters (interaction potential, size, shape, mixtures of different kinds of particles) and by imposing external conditions like confinement, shear, compression and temperature \cite{16}. A thorough understanding of these processes is necessary for the design of methods to create nanostructures utilizing the self-assembly of particles. The aim here is to be able to predict which pattern will form under which conditions.

Several studies using Monte Carlo computer simulations have already been carried out investigating the structure formation in colloidal crystals. Some of them have focussed on the effect of different kinds of walls \cite{17,18}, on the impact of external fields \cite{19} and on confinement incommensurate with the crystalline structure leading to the formation of a soliton staircase \cite{20,21,22} in a one-component colloidal system crystallizing in a triangular lattice structure. Additional studies have investigated the melting \cite{23}, the deformation \cite{24}, the correlation between channel geometry and pattern formation \cite{25,26} and frustration \cite{27} of such two-dimensional systems of purely repulsive particles under confinement. Others have considered binary mixtures which crystallize into a square lattice structure \cite{Stefan_Dipl,Stefan_Dipl_Paper,32,33,34,35,36} and fluid systems \cite{70}. In this paper we will concentrate on the influence of shear on a one-component colloidal crystal with a hexagonal lattice structure with and without misfit. The aim of our work is to clarify the range over which moving boundaries induce flow of particles in such a soft crystal, and to study the extent to which such nonequilibrium phenomena can still be described by local equilibrium concepts.

In the following section, we will describe the model system and the simulation parameters. In section \ref{sec_results} we will report on the results of our simulations with sheared walls for the case where the effective periodic wall potential is commensurate with the lattice period of the crystal. We will present snapshots of the created configurations, velocity profiles and angular distributions. Then we will move to the results for the case where the lattice period is no longer commensurate with the wall potential and present the same quantities in order to compare them. As a last step, we will look at the mean square displacement of the particles in both cases. In section \ref{sec_conclude} we will briefly summarize the main conclusions of this work.

\section{Details of the simulation}

\begin{figure}[htbp]
\centering
\includegraphics[scale=0.29, clip=true]{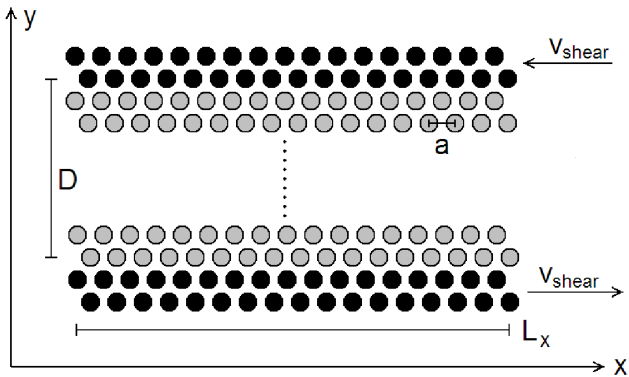}
\caption{Sketch of the system geometry with the directions of the axes and the sheared wall particles. Black dots denote wall particles, grey dots freely movable particles.}
\label{Skizze_geometrie}
\end{figure}

In order to investigate the influence of shear in combination with confinement in a two-dimensional colloidal crystal, we set up a model system consisting of $N=73440$ particles, arranged in a slit geometry which is elongated along the x-axis. In the x-direction, periodic boundary conditions are applied, while the system is confined in the y-direction by two walls consisting of two rows of frozen particles each (see fig.~\ref{Skizze_geometrie}. These ``frozen'' particles interact with the ``real'' particles via the same potential with which the particles interact with each other. We use 30 rows parallel to the walls (in a commensurate arrangement with the walls) or only 29 rows (which then are incommensurate due to higher particle densities in the resulting smaller ``volume'').

Shear is implemented by moving these structured walls in the x-direction with a certain, constant velocity. The two walls on either side of the stripe are moved with the same velocity, but in opposite directions.

The particle interaction is defined via a purely repulsive pair potential like the potential which would be felt by magnetic particles when a magnetic field is turned on. However, the resulting potential is inconvenient for computer simulations due to its slow $r^{-3}$ decay with distance, and since there are not yet any experiments for sheared colloid crystals to compare with, we prefer to use a computationally more efficient $r^{-12}$-potential with a cutoff (in order to make it strictly short-ranged), a shift (in order to ensure that it ends with a value of zero), and a smoothing factor (to make it differentiable). A further motivation to use this model potential is that it is essentially the same potential used in the studies of confinement effects on colloidal crystals without shear \cite{17,18,19,20,21,22} and of two-dimensional melting \cite{82}. This potential is defined as \cite{Stefan_Dipl,Stefan_Dipl_Paper}:
\begin{equation} \label{eq_pot}
V(r) = \left[\epsilon \left(\frac{\sigma}{r}\right)^{12} - \epsilon \left(\frac{\sigma}{2.5}\right)^{12}\right] \cdot \left[\frac{(r-2.5\sigma)^4} {h^4 + (r-2.5\sigma)^4}\right]
\end{equation}
where $h=0.01 \sigma$ and $\sigma=1$ were chosen. Thus the particle diameter $\sigma$ is our unit of length, and $\epsilon=1$ sets the scale for the temperature.

The particles are arranged in a hexagonal lattice structure. In the case without misfit, the walls are placed such a distance apart, that a perfect crystalline lattice structure with $n_y=30$ rows of $n_x=2160$ particles fits between them. The dimensions of the system $L_x$ and $D$ follow from the lattice constant $a$ via $L_x=n_x a$ and $D=n_y a \sqrt{3}/2$. In the case with a misfit, the walls are placed closer together, thus shrinking the width of the stripe to $D=(n_y-\Delta) a \sqrt{3}/2$. This parameter $\Delta$ is referred to as ``misfit'' in the following.

\begin{figure}[htbp]
\centering
\includegraphics[scale=0.1, clip=true]{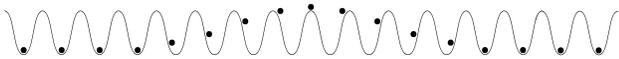}
\caption{Explanation of solitons: As there are more particles in the the inner rows, the distance between these particles is on average smaller than the distance between the particles in the outer rows forming the rigid walls. These rigid rows act like a given external periodic potential on the mobile particles. Therefore the particles in the inner rows are no longer placed at the minima of the potential created by the particles in the outer rows, thus creating a soliton.}
\label{Skizze_solitonen}
\end{figure}

At larger misfits, the structure of the crystalline strip is rearranged to 29 rows as has been investigated in previous studies by Chui et al \cite{20,21,22}. At a misfit of $\Delta=2.2$ we therefore started a part of our simulations with a compressed 30-row-structure, which quickly rearranged itself into a 29-row-structure upon shearing, and a part of our simulations were started with a 29-row-structure. This structure with 29 rows contained as many particles as the structure with 30 rows. Therefore, those particles that had been in the 30th row had to be distributed amongst the remaining 29 rows. As the number of particles in the walls did not change, previous studies showed, that the rows directly adjacent to the walls remain free of extra particles while all of the inner rows incorporate the same number of extra particles thus creating a lattice with a smaller lattice constant $a_x$ in x-direction. But due to the misfit, the lattice constant in y-direction is smaller, too, which stabilized this new crystalline structure. Nevertheless, the incommensurability of the crystalline structures of the inner rows with the rows directly adjacent to the walls leads to the formation of a soliton staircase \cite{20}. Fig.~\ref{Skizze_solitonen} explains how solitons are created by the incommensurability of two rows with a different number of particles.

\begin{figure}[htbp]
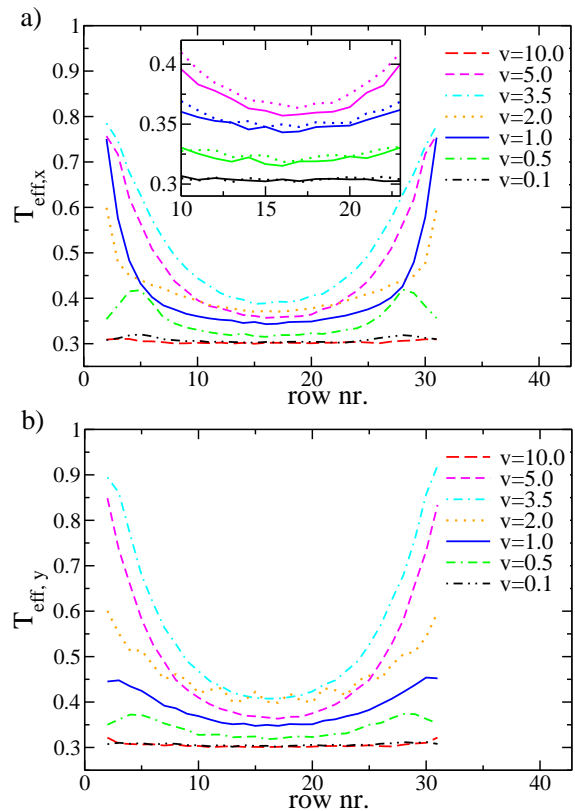

\centering
\includegraphics[scale=0.3, clip=true]{fig3a.eps}\\
\includegraphics[scale=0.3, clip=true]{fig3b.eps}
\caption{Effective temperature, calculated from the velocity distribution in x-direction (a) and y-direction (b) by fitting it via $f(v_x)=a_0 \cdot e^{(v_x-\bar v_x)^2/(2 T_{eff})}$ (and likewise for $v_y$, with $\bar v_y=0$). Shear velocities of the walls as indicated. Starting configuration with 30 rows without misfit ($\Delta=0$), at $T=0.3$. The inset in part (a) compares the effective temperatures  $T_{eff,x}$ (solid lines) and $T_{eff,y}$ (dotted lines) for $v=5.0$, $v=1.0$, $v=0.5$ and $v=0.1$ (from top to bottom).}
\label{Graph_Temp_eff}
\end{figure}

The simulations were carried out using the program package HOOMD-blue \cite{80,81}, which is designed to run on graphic cards and provides the option to use Langevin Dynamics by only defining a parameter $\gamma$ for the drag of the particles, a timestep $\Delta t$ and a temperature $T$. It uses the well-known Velocity Verlet integration method and adds a force $\vec{F}=-\gamma \vec v + \vec{F}_{rand}$ to the force exerted on each particle by the interaction with its neighbouring particles \cite{Allen,Evans}. Here, $\vec v$ is the particle's velocity and $\vec{F}_{rand}$ is a random force with a magnitude chosen via the fluctuation-dissipation theorem to be consistent with the specified drag $\gamma$ and temperature $T$. In all of the simulations presented here, we used $\gamma=0.5$. Note that our particles have mass $m=1$, and time $t$ is measured in the standard Molecular Dynamics time unit $\tau=t\sqrt{\frac{\epsilon}{m \sigma^2}}$. Because of the huge speed-up that occurs when MD simulations are run on graphic cards we initially used them as well, but experienced problems with the accumulation of systematic errors. However, as the graphic card code uses single instead of double precision, we had to carry out the simulations on regular CPUs as the higher accuracy turned out to be necessary for our crystalline systems. We chose a timestep of $\Delta t=0.002$ and several temperatures as indicated in the respective paragraphs. All of these temperatures were below the melting transition which would occur for this system at the chosen density of $\rho=1.05$ (for the case without misfit) at about $T=1.35$ \cite{20}. Nevertheless, the Langevin thermostat only takes into account the velocities of the particles in the bulk (and not the velocity of the wall particles which are moved with constant velocity in order to introduce shear into the system). This leads to a higher effective temperature of the particles due to their interaction with the moving wall particles, because the wall particles are constantly introducing friction and thereby increasing the temperature (see fig.~\ref{Graph_Temp_eff}). The Langevin dynamics are aiming at restalling the initial temperature, but as the wall particles move again with every new time step, more heat is added to the system, and an equilibrium between the cooling due to the Langevin thermostat and the heating due to the sheared walls is reached. This equilibrium temperature is higher than the temperature which was set in the Langevin thermostat. But this temperature increase due to the sheared walls does not only depend on the velocity of the walls, but also on the interactions of the particles with the walls. So at rather low shear velocities between $v=0.1$ and $v=3.5$, the effective temperature of the system increases with increasing shear velocity, but for a very high shear velocity of $v=50.0$, the effect of the shearing on the temperature is almost negligibly small again, because such a fast shearing has almost no effect on the particles next to it, which feel only an averaged-out wall potential in this case.

Since near the walls the effective temperature extracted from the $x$ and $y$-components of the velocities of the particles disagree, it is clear that the ``fluid layers'' near the moving walls are completely out of equilibrium, since the concept of a ``local temperature'' which is anisotropic does not make much sense. However, for small velocities the ``temperature'' profiles $T_{eff,x}(y)$ and $T_{eff,y}(y)$ agree within statistical errors, and the concept of a ``local temperature'' in the slit interior can be used. Then the observed profile of this local temperature $T_{eff}(y)$ can be interpreted as follows: due to the large friction between the fluid-like layers and the moving walls, these layers act like a source where heat is constantly pumped into the system. This heat is transported via conduction into the interior of the slit (due to the action of the thermostat, heat is removed everywhere in the slit, and a steady-state temperature profile is established). Motivated by corresponding solutions of the heat conduction equation, we can fit these local temperature profiles by $T(y)=A_0[exp(-z/\lambda)+exp(-(D-z)/\lambda)]+A1$, where $A_0$, $A_1$ and $\lambda$ are phenomenological parameters that depend on $v$. Unfortunately, due to the large fluctuations the fitting of three parameters is somewhat
uncertain and hence we do not dwell on the velocity dependence of these parameters. It would be interesting to study how this dependence changes when the friction constant is varied, but a study of this problem would reqiure very large computational resources and hence is beyond the scope of the present paper.

\section{Results}
\label{sec_results}

\begin{figure}[htbp]
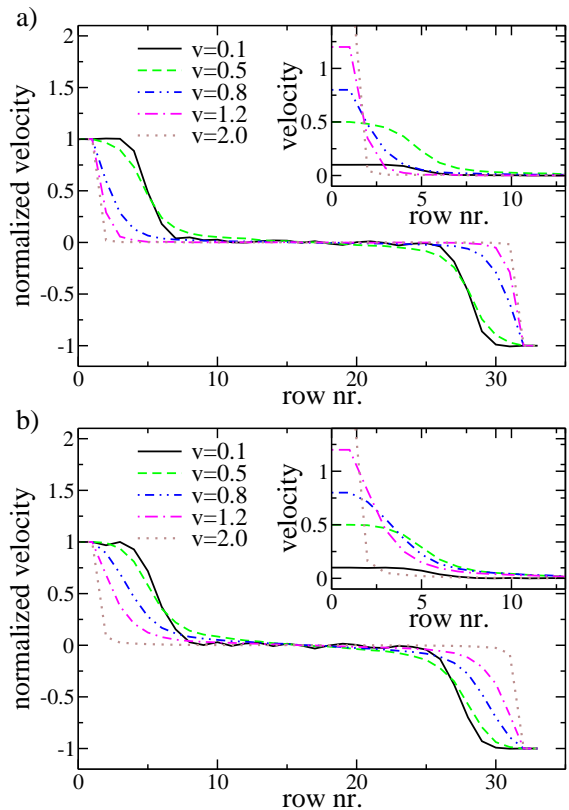

\centering
\includegraphics[scale=0.3, clip=true]{fig4a.eps}\\
\includegraphics[scale=0.3, clip=true]{fig4b.eps}
\caption{Velocity profiles. Shear velocities of the walls as indicated. Starting configuration with 30 rows without misfit ($\Delta=0$), for two temperatures $T=0.3$ (a) and $T=0.5$ (b).}
\label{Graph_vel_profiles_30_000}
\end{figure}

Without a misfit (i.e. $\Delta=0$), where the crystal consists of 30 rows, slow shearing of the walls has a relatively large effect on the velocities of the particles. Fig.~\ref{Graph_vel_profiles_30_000} shows velocity profiles obtained by sorting the particles into rows depending on their position in the y-direction and then calculating the average particle velocity for each row of particles. As the maximum average particle velocity is given by the shear velocity of the walls, we normalized the velocity profiles by the shear velocity. But we also show the non-normalized velocities in an inset of the graphs (fig.~\ref{Graph_vel_profiles_30_000}) as a higher absolute velocity will lead to more disorder in the crystal, while the normalized velocities indicate how strong the effect of the shearing is. 

At a temperature of $T=0.3$ and for the case of $\Delta=0$ this results in velocity profiles (fig.~\ref{Graph_vel_profiles_30_000}a) where the two smallest shear velocities ($v=0.1$ and $v=0.3$) lead to the (relatively) largest particle velocity in the first $4-5$ rows closest to the walls. This means that slow shearing of the walls has the strongest effect on pulling the particles along with the walls. Of course, if one compares the absolute, non-normalized velocities inside the crystal, these velocities are larger at a shear velocity of $v=0.3$ and $v=0.5$, as particles next to the walls obtain a higher velocity due to the faster walls, even if a smaller percentage of the walls' velocity is transferred to the particles next to them. The higher absolute velocity also leads to more disorder inside the crystal, which then leads to a stronger influence of the shearing even on particles in the inner rows of the crystal. Thus, there is clear evidence for shear band formation, as will be discussed below.

\begin{figure}[htbp]
\centering
\includegraphics[scale=0.3, clip=true]{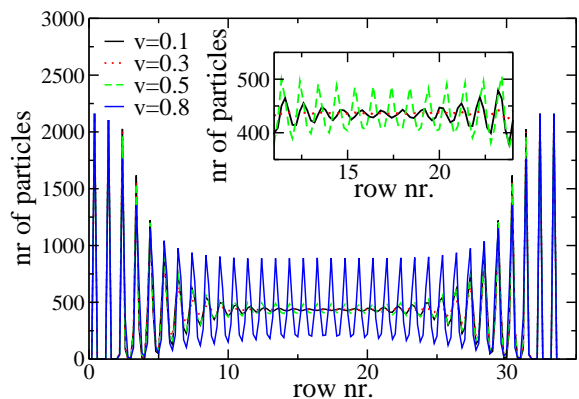}
\caption{Density profiles. Simulations were carried out at a temperature of $T=0.3$ for a system with 30 rows without a misfit ($\Delta=0$). Shear velocities of the walls as indicated. The strongly oscillating profile represents $v=0.8$ and was omitted in the inset, where a magnified plot of the central part of the slit is shown. The normalization of the density profile $\rho(y)$ is given in terms of the total number of particles, $N=\int_0^D \rho(y) dy$.}
\label{Graph_density_T03}
\end{figure}

\begin{figure}[htbp]
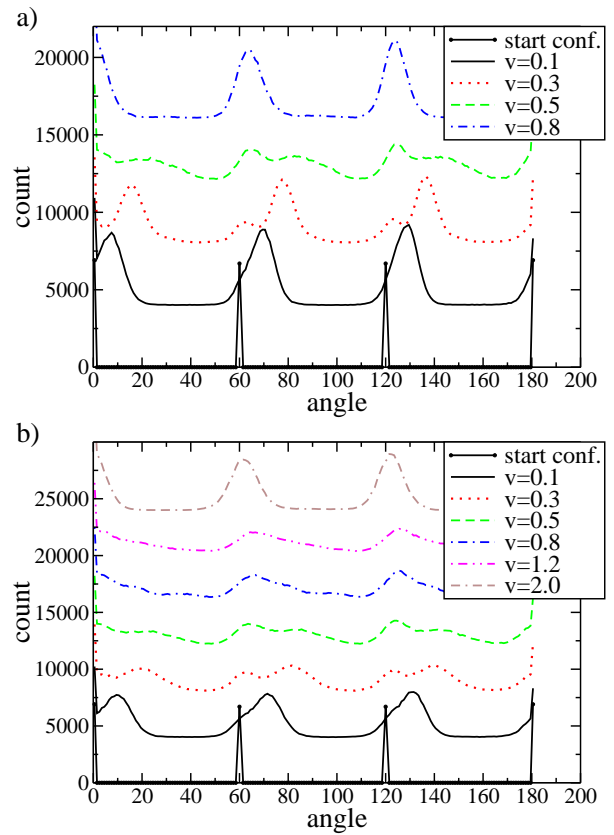

\centering
\includegraphics[scale=0.3, clip=true]{fig7a.eps}\\
\includegraphics[scale=0.3, clip=true]{fig7b.eps}
\caption{Distributions of angles between the line connecting each pair of two neighbouring particles and the x-axis. Shear velocities of the walls as indicated. The graphs for different shear velocities are shifted vertically by $4000$ in order to give a better overview. Starting configuration with 30 rows without misfit ($\Delta=0$), temperature $T=0.3$ (a) and $T=0.5$ (b).}
\label{Graph_angular_distrib_30_000}
\end{figure}

This disorder is also reflected in the density profiles (fig.~\ref{Graph_density_T03}) and in the angular distributions (fig.~\ref{Graph_angular_distrib_30_000}a). The density profiles show clear peaks indicating a crystalline structure for higher shear rates, but at a shear rate of $v=0.3$, the peaks corresponding to the inner rows of the crystal are almost non-existent any more. This can be explained by the changes in the crystalline structure discussed below.

\begin{figure}[htbp]
\centering
\includegraphics[scale=0.132, clip=true]{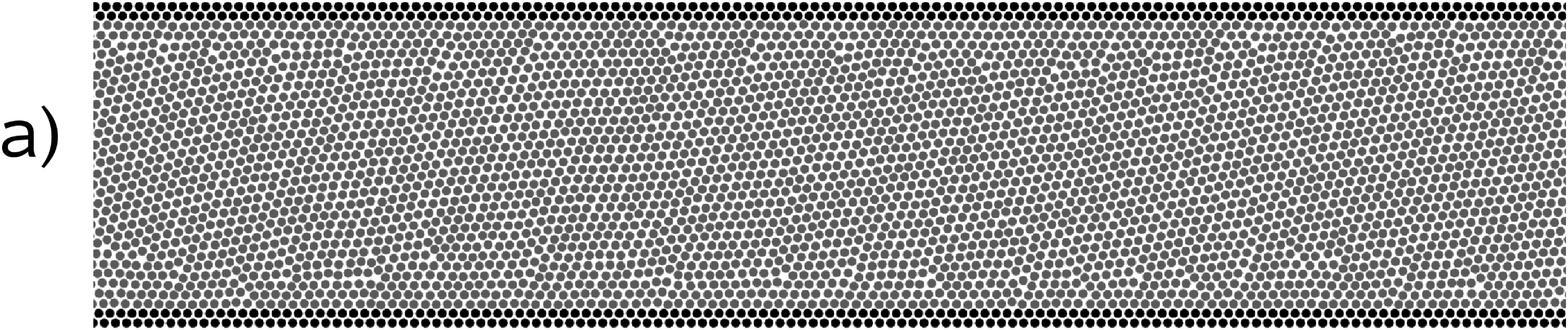}\\
\includegraphics[scale=0.132, clip=true]{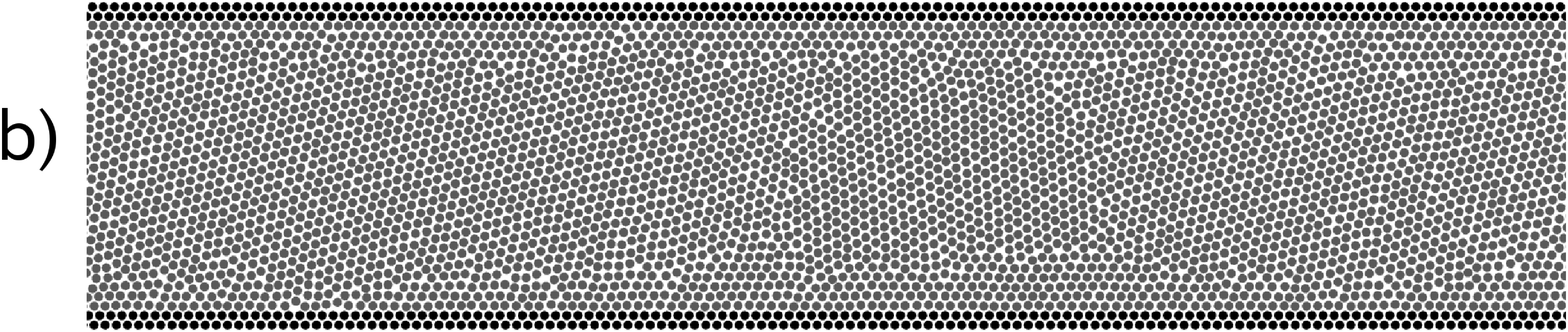}\\
\includegraphics[scale=0.132, clip=true]{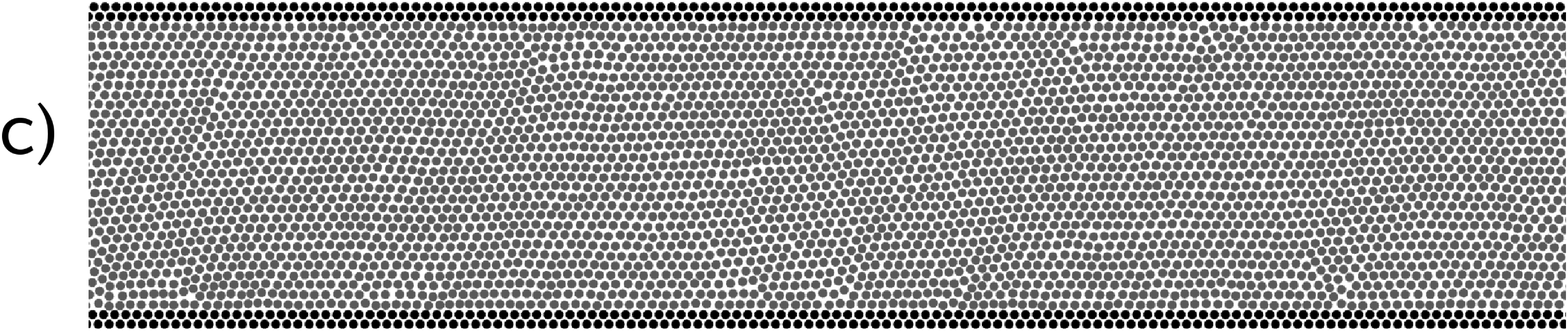}
\caption{Snapshots of a section of a system with sheared walls. The starting configuration contained 30 rows and the simulation was carried out at a temperature $T=0.3$ and without a misfit ($\Delta=0$). The particles in the two first rows on top and the two last rows at the bottom (that are at rigidly fixed positions relative to one another and contribute the moving walls) are highlighted by black dots while the mobile particles are shown as grey dots. Shear velocities of the walls were chosen as $v=0.1$ (a), $v=0.3$ (b) and $v=0.8$ (c)}
\label{Graph_snapshot_shearing}
\end{figure}

\begin{figure}[htbp]
\centering
\includegraphics[scale=0.3, clip=true]{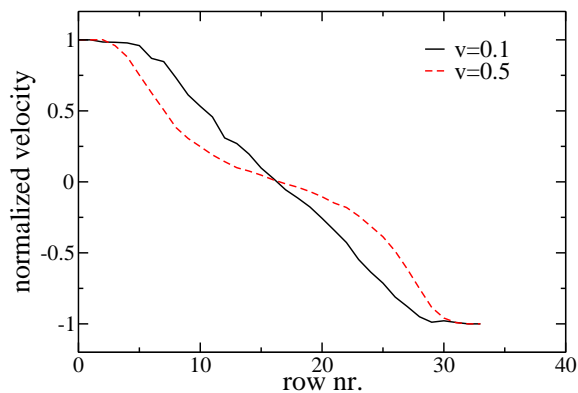}
\caption{Velocity profiles for $T=1.5$. Shear velocities of the walls as indicated. Starting configuration with 30 rows without misfit ($\Delta=0$).}
\label{Graph_linear_vel_profiles}
\end{figure}

Obviously the profiles of the velocity never resemble the (approximately) linear velocity profiles representing simple Couette flow, that one encounters when one shears simple \cite{90} or complex \cite{91,92} fluids. This is due to the fact that we are shearing a crystalline solid here. At higher temperatures and small shear velocities almost linear velocity profiles can be obtained for our system as well (see fig.~\ref{Graph_linear_vel_profiles}).

\begin{figure}[htbp]
\centering
\includegraphics[scale=0.3, clip=true]{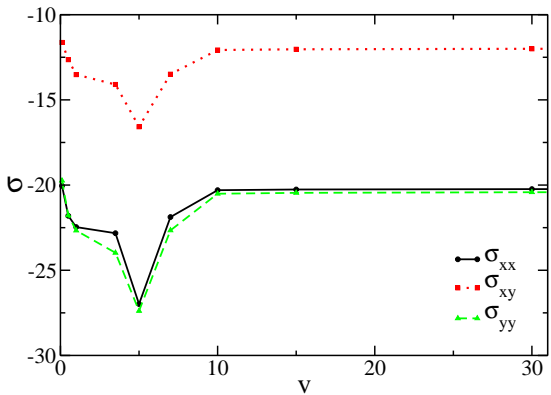}
\caption{Flow curve for $T=0.3$ and starting configurations with 29 rows at a misfit of $\Delta=2.2$. $\sigma_{xx}$, $\sigma_{xy}$ and $\sigma_{yy}$ denote the respective components of the stress tensor, while $v$ indicates the velocity with which the walls are being moved.}
\label{Graph_flow_curve_dip}
\end{figure}

The shearing of the walls leads (at certain shear rates) to a tilting of the crystalline layers and even to breaks in the crystalline structure, where parts of the crystal are oriented in a different direction from the rest. This effect is illustrated in the snapshots shown in fig.~\ref{Graph_snapshot_shearing}. In order to analyze it quantitatively, we computed for each pair of two neighbouring particles the angle between the line connecting these two particles and the x-axis. In a perfect hexagonal lattice without misfits and without defects, this would lead to peaks at $0^{\circ}$, $60^{\circ}$, $120^{\circ}$ and $180^{\circ}$ (we did not compute any angles larger than $180^{\circ}$).

At a shear velocity of $v=0.1$, the peaks are centered around slightly larger angles due to a tilting of the crystalline layers which is induced by the shearing (fig.~\ref{Graph_snapshot_shearing}a and fig.~\ref{Graph_angular_distrib_30_000}). At shear velocities of $v=0.3$, $v=0.5$ and (less pronounced) of $v=0.8$ the angular distribution has two peaks as the crystal breaks into several pieces and generally only the layers close to the walls are aligned with the walls (creating a peak in the angular distribution close to $60^{\circ}$, $120^{\circ}$ etc), while the crystalline parts inside of the crystal prefer a different orientation, thus creating the second peak in the angular distribution (fig.~\ref{Graph_snapshot_shearing}b and fig.~\ref{Graph_angular_distrib_30_000}). These two peaks are very distinct at $v=0.3$, while at $v=0.5$ there is more disorder inside of the crystal due to the higher shear velocity. Starting from $v=0.8$, this second peak disappears again as the walls are moving too fast to have much influence on the crystalline structure any more. This is due to the fact that the fast shearing of the walls leads to an averaged out wall potential which the particles next to the wall feel, which is not able to drag particles along with the walls any more.

It is interesting to note that the above results may be understood as follows. Firstly, at low velocities, the velocity profile shown in fig.~\ref{Graph_vel_profiles_30_000} may be interpreted as being composed of three prominent shear bands \cite{93}, a feature of a typical non-Newtonian, shear-thinning material. The two fast moving bands lie close to the boundaries and flow in opposite directions sandwiching a central immobile region. In some respects our flowing solid therefore behaves like a moving paste, or soft-glassy matter \cite{94} in a fashion which is complementary to that seen in the experiments of Coussot et al. \cite{95} where a central mobile region is flanked on either side by immobile shear bands adjacent to the boundaries. As the shear rate is increased, the width of the mobile shear bands initially increases as expected. In order to verify that we indeed do have shear banding in our system, we plot the average stress as a function of shear rate in fig.~\ref{Graph_flow_curve_dip}. The resulting flow curve shows non-monotonic behavior typical of shear thinning systems. However, surprisingly, instead of obtaining a linear Newtonian region at high shear rates,  the stress stabilizes to a constant signifying the presence of a shear band of vanishingly small viscosity. Referring back to the calculated velocity profiles, (see fig.~\ref{Graph_vel_profiles_30_000}) we observe that the width of the shear bands decreases with increasing velocity in this regime, so that at large velocities (shear rate), we obtain thin, highly mobile boundary layers near the walls while most of the solid is immobile. For such velocities, the solid re-crystallizes.

Further, in order to characterize the behavior of the solid at high wall velocites, we refer to the plot of the anisotropic, local, effective temperature (obtained by fitting a Gaussian to the probability distribution of the $x$ and $y$ components of the velocity) already shown in fig.~\ref{Graph_Temp_eff}. We observe that both these effective temperatures show non-monotonic behavior, initially increasing and then decreasing as the velocity increases leading to re-crystallization. Regaining crystalline order with high drive velocities is also seen in crystals driven over random pinning potentials \cite{96}. This implies that our system {\it never} shows a Newtonian regime at any flow velocity if the temperature is small enough to obtain crystallization. At higher temperatures, when the solid melts in equilibrium, we do obtain (nearly) Newtonian flow (fig.~\ref{Graph_linear_vel_profiles}).

At higher temperatures, but still below the melting temperature, it is energetically easier for the crystal to split up into several domains, which can then be tilted with respect to the walls and can be oriented in different directions. Therefore the velocity profiles at temperature $T=0.5$ have a more pronounced shape even at higher shear velocities (see fig.~\ref{Graph_vel_profiles_30_000}b). The same effect is visible in the angular distributions (fig.~\ref{Graph_angular_distrib_30_000}b): The ``second peak'' in the angular distributions only disappears at considerably higher shear rates than in the case of $T=0.3$. Apart from this, all shear rates at $T=0.5$ lead to higher average particle velocities than in the case of $T=0.3$ as the higher temperature makes it easier for particles to move along with the walls.

A different picture arises if one introduces a misfit into these simulations. As described above, a misfit means that the structured walls are placed closer together. In simulations without shearing, this would lead to the creation of a soliton staircase \cite{20} (compare fig. \ref{Skizze_solitonen}). It is also already known that there is a significant hysteresis in the transition between structures with 30 rows and structures with 29 rows with the same overall number of particles. These structures with 29 rows are the energetically preferred configuration for misfits greater than about $\Delta=1.7$, but if a simulation is started with a structure of 30 rows, it will not spontaneously change its structure to 29 rows (in simulations without shear) until about a misfit of $\Delta=2.0$. But even at larger misfits, simulations without shear have to be equilibrated for a certain time until this transition takes place. This transition happens much faster when sheared walls are introduced into the system. For example, for a system starting with 30 rows at a misfit of $\Delta=2.2$ and at a temperature of $T=0.3$, the system still consisted of 30 rows after 1 mio. time steps in the case without sheared walls, but had already 29 rows after just 20000 time steps in the case of a simulation where the walls were sheared with a velocity of $v=0.5$. As will be discussed later on, at this shear velocity the structure breaks and changes again if the simulation run is continued.

In the following, we will concentrate on simulations at a misfit of $\Delta=2.2$, which leads to a structure with 29 rows independently of whether the starting configuration contained 30 or 29 rows. It is interesting to investigate the structures which develop here as they do not contain solitons any more due to the shearing of the walls. Instead, it is now possible for the system to re-arrange its particles continually in a way that no particles have to sit on the maxima of the potentials, as there is enough disorder in the system, especially close to the walls.

\begin{figure}[htbp]
\centering
\includegraphics[scale=0.3, clip=true]{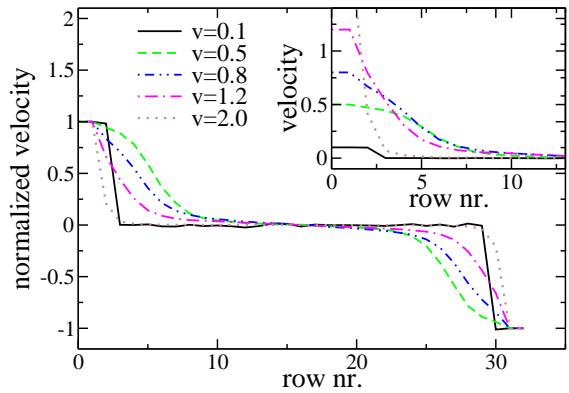}
\caption{Velocity profiles. Shear velocities of the walls as indicated. Temperature $T=0.3$. Starting configuration with 29 rows at a misfit of $\Delta=2.2$.}
\label{Graph_vel_profiles_29_30_220}
\end{figure}

There are several mechanisms of dealing with the different number of particles in the different rows which lead to the formation of solitons in the case without shear. At low shear rates and if the starting configuration already contains 29 rows, the row next to the walls (which contains the same number of particles as the wall rows, but less than all of the inner rows), moves along with the walls, while the inner rows remain almost without movement (fig.~\ref{Graph_vel_profiles_29_30_220}, curves for $v=0.1$ and $v=0.3$). This would in principle lead to solitons which are moved along with the walls as the rows directly adjacent to the walls still do not have the same crystalline structure like the inner rows. But as these rows are moving, the particles they contain are not always at the ideal lattice positions and therefore the solitons are strongly ``smeared out''.

If one starts with a structure of 30 rows at the same shear rate, the structure changes into 29 rows. Apart from equilibration effects, we obtained the same results as in the case where we started with a well-defined 29-row-structure.

\begin{figure}[htbp]
\centering
\includegraphics[scale=0.3, clip=true]{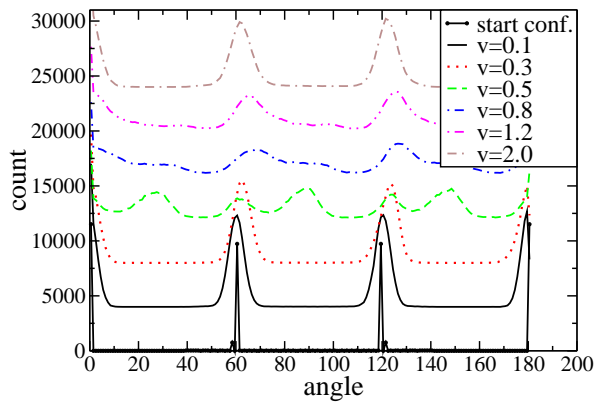}
\caption{Angular distributions. Shear velocities of the walls as indicated. The graphs for different shear velocities are shifted vertically by $4000$ in order to give a better overview. Starting configuration with 29 rows at a misfit of $\Delta=2.2$. Temperature $T=0.3$.}
\label{Graph_angular_distrib_29_220}
\end{figure}

At slightly higher shear velocities ($v=0.5$ and, less pronounced, also $v=0.8$), the same effect as in the case without misfit is observed: The crystalline layers are tilted and the inner part of the crystal breaks off from the rest and changes its orientation, thus creating a rather strong flow in the layers close to the wall and less flow inside of those parts of the crystal, which have turned around. In this case, the solitons are a part of the general disorder and therefore not visible in any way. The change of the structures becomes visible in the angular distributions shown in fig.~\ref{Graph_angular_distrib_29_220}.

Just like in the case without misfit, a further increase of the shear velocity leads to a less pronounced effect of the shearing again (see curves for $v=1.0$ up to $v=3.5$) and the crystalline structure resumes a greater order. Still, due to the shear there is enough disorder in the rows close to the walls that there are no distinguishable solitons.

Here, it is interesting to study the case of very high shear velocities ($v=50.0$) as the shearing seems to stabilize the structure with 30 rows. So, if the starting configuration contains 30 rows which contain the same number of particles in each row and therefore no solitons, this structure is stabilized by the fast moving walls.

\begin{figure}[htbp]
\centering
\includegraphics[scale=0.14, clip=true]{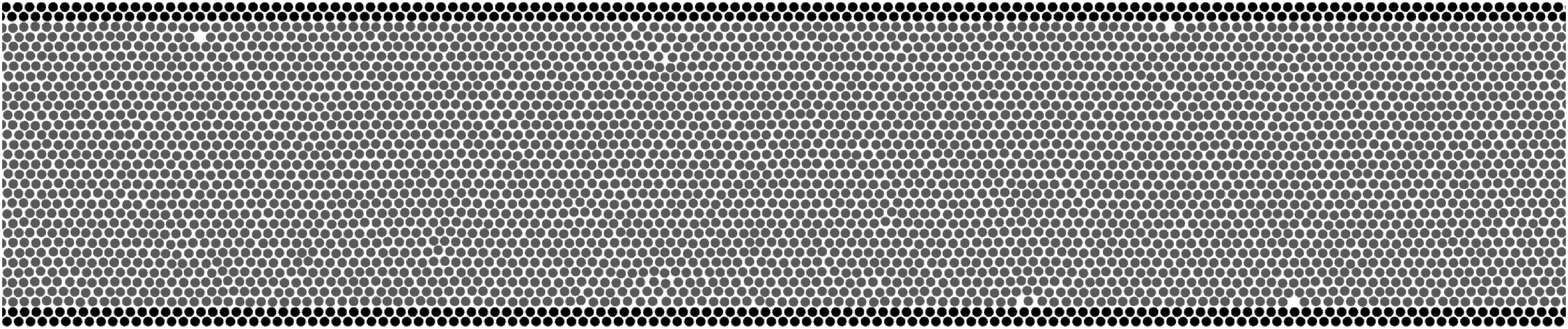}
\caption{Snapshot of a section of a system with sheared walls. The starting configuration contained 29 rows and the simulation was carried out at a temperature $T=0.3$ and a misfit of $\Delta=2.2$. The high shear velocity of the walls ($v=50.0$) leads to the creation of holes.}
\label{Graph_snapshot_shearing_50}
\end{figure}

If the starting configuration contains 29 rows and the appropriate number of extra particles in the inner rows, a different effect can be observed: As the structured walls are moving very fast, the rows directly adjacent to the walls only feel an averaged-out wall potential and are thus not stabilized in a structure with the same number of particles as the walls any longer. Therefore it becomes energetically preferrable for these rows to contain the same number of particles like the inner rows, which contain extra particles. But as the number of particles in the crystal is constant, there are no particles to fill up the outer rows. Due to this, holes are created in the outer rows which take the role of additional particles. These holes can even diffuse further inside the crystal. A snapshot showing this effect is presented in fig. \ref{Graph_snapshot_shearing_50}.

Re-running all of these simulations at the same misfit and a higher temperature ($T=0.5$) leads to results that can be expected from the case without misfit: As there is generally more disorder in the system if the temperature is higher, the velocity profiles become more pronounced for higher shear rates. In addition, the crystal starts to split up into several domains and to change its orientation at lower shear rates already than in the case of $T=0.3$.

\begin{figure}[htbp]
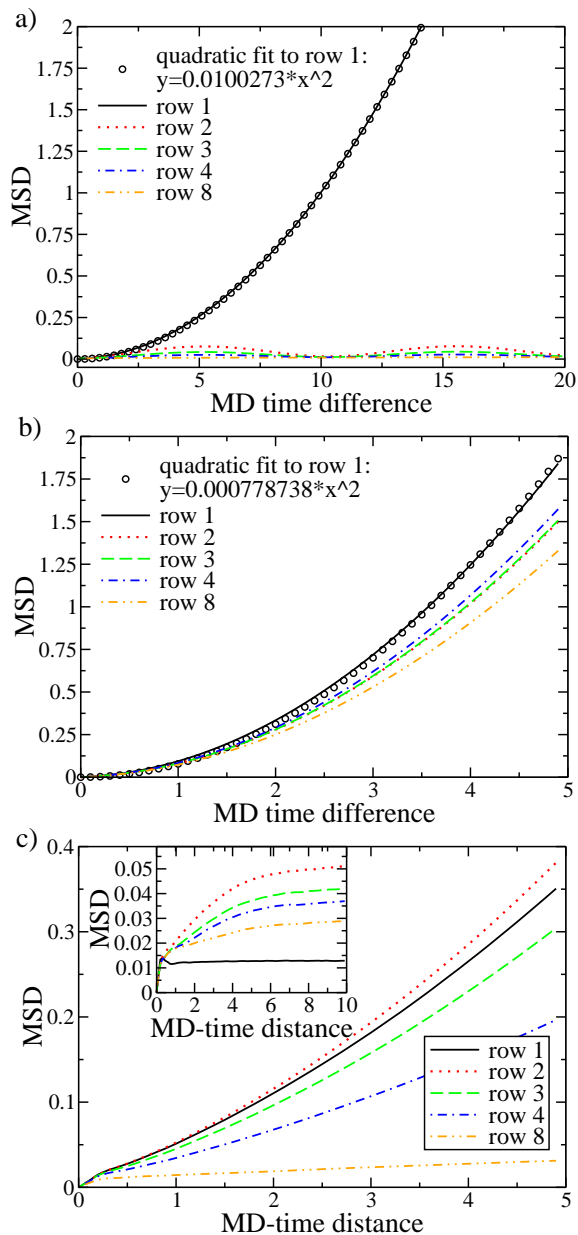

\centering
\includegraphics[scale=0.3, clip=true]{fig13a.eps}\\
\includegraphics[scale=0.3, clip=true]{fig13b.eps}\\
\includegraphics[scale=0.3, clip=true]{fig13c.eps}
\caption{MSD (in x-direction) of the particles in different rows. The starting configuration contained 29 rows in every case and the simulation was carried out at a temperature $T=0.3$ and a misfit of $\Delta=2.2$. (a) The shear velocity of the walls were chosen as $v=0.1$. The oscillation is due to the movement of the wall particles which leads to an oscillating wall potential for the particles next to the walls. (b) Shear velocity $v=0.5$, (c) $v=1.2$. The inset shows the MSD in the case without shearing for comparison.}
\label{Graph_shearing_MSD}
\end{figure}

Another interesting property of these systems is the time-dependence of the mean square displacement (MSD) of the particles. Here the MSD is defined as $MSD=<[x_i(t+t_0)-x_i(t_0)]^2>$, where the average $<...>$ includes an average over all particles in a row and over the initial time $t_0$. We are only presenting results for the MSD in the x-direction parallel to the walls here.

As described above, in cases with a misfit of $\Delta=2.2$ and at low shear velocities ($v=0.1$), only the outermost row of particles moves along with the walls. This leads to an MSD as shown in fig.~\ref{Graph_shearing_MSD}, where the MSD of the particles in the first row is an almost perfect quadratic curve of the form $MSD=a\cdot t^2$. As the MSD is the quadratic displacement, this means that the fit parameter $a=v^2$, $v$ being the velocity of the particles in this row. As one can read off from the graph, the velocity of the particles in this case is almost exactly the velocity with which the walls are being sheared. The MSD of the rows next to the outermost one have significantly smaller values and a periodic shape as the particles in these rows are trying to cope with the moving walls. The MSD of the inner rows is almost constant and independent of the walls, which means that the particles just move around randomly.

At a shear velocity of $v=0.5$ (fig.~\ref{Graph_shearing_MSD} middle), where the orientation of all the inner parts of the crystal changes, the MSD looks completely different: The particles in the first row are still moving more than the ones in the inner rows, but as the particles of all rows are mixing in the course of the simulation, the MSDs of different rows are getting more similar the longer the simulation runs, until finally the particles of all rows are equally distributed in the crystal and the MSD of particles stemming from different rows can not be distinguished any more. Besides, all values of the MSDs are very large here (note the different scales of the y-axes in fig.~\ref{Graph_shearing_MSD}).

At even higher shear velocities, where no parts of the crystal change their orientations any more, the values of the MSD are generally smaller again. As illustrated in fig.~\ref{Graph_shearing_MSD} (c), the MSD in the row directly adjacent to the walls is the second largest one and can not be fitted with a quadratic curve of the form $MSD=a\cdot t^2$ any more. This is due to the fact that the particles' movement is not primarily the movement along with the walls any more. The MSD in the row next to this row is larger. This is due to the incommensurability of the different lattice spacings. The row directly adjacent to the walls originally contained as many particles as the walls while the inner rows contain extra particles due to the disappearance of one row because of the misfit. In the rows further inside of the crystal, the MSD is smaller as the effect of the shear and of the incommensurabilities becomes smaller.

If one compares the MSD of these simulations under shear with simulations without shear at the same misfit, the effect of the solitons becomes visible (see fig.~\ref{Graph_shearing_MSD}, inset of graph c): In the case without shear, the particles in the row directly adjacent to the wall do not move much as their structure is commensurate with the walls and the walls are not moving. Therefore their MSD is the smallest one and even shows the overshoot, which one would expect to see in a crystal with harmonic potentials where the particles are swinging around their ideal lattice positions with a certain periodicity. The MSD of the rows adjacent to this one are significantly larger in this case as the solitons are located in these rows, which causes the particles to move around more. In the inner rows, the MSD becomes smaller again as there are no solitons in these rows. This is clearly different from the values of the MSD in the case with shear, where no clearly definable solitons are created and therefore all of the outer rows exhibit comparable values of the MSD as the incommensurabilities are shared between these rows and lead to a lot of movement. Besides this, the shape of the curves is different and the absolute values of the MSD are larger as the shearing leads to a small, but detectable movement along with the sheared walls, while in the case without shear no direction is preferred.

The assignment of particles to rows is done according to the initial position of each particle at the beginning of the time interval $\Delta t$ used as abscissa variable here since in the course of time particles change their rows.

\section{Concluding remarks}
\label{sec_conclude}

In this paper we discussed the influence of shear on the structure of a two-dimensional colloidal crystal in confinement between parallel walls. We observed that in the case without misfit, at slow shear velocities the layers of the crystal are tilted, while the crystal breaks into several pieces with different orientations at medium shear velocities and remains almost undisturbed at high shear velocities. We showed that the distribution of angles inside of the crystal is an important quantity and gives a good insight into the structure of the crystal, along with velocity profiles.

In the case with a misfit, however, the behaviour is slightly different, especially at low and at high shear rates. At low shear rates, only one single row of particles moves along with the walls and at high shear rates holes are created in the structure. But the breaking up into several pieces also occurs at slighly different shear velocities. This is due to the incommensurability in the case with a misfit, which would lead to the creation of solitons in the case without a misfit. Obviously, the shearing of the crystal changes this behaviour significantly.

Clearly, our results are somewhat qualitative. One could in principle run more simulations at different temperatures and different shear velocities in order to investigate the occuring structural transitions (e.g. from 30 to 29 rows and order-disorder-transitions) with higher precision. But this does not seem very sensible at this moment because there is no theory with which the results could be compared and because experiments with such colloidal systems are possible \cite{1,2,3,4,5,6,7,8,9,10,11,14,34}, but typically use slightly different parameters for the particle interactions, the box sizes and the walls. Therefore a rather qualitative study makes sense in this particular case and helps to understand the complex interplay of confinement and shear and its influence on the structure formation of colloidal crystals.

\section{Acknowledgements}

This work was performed in the framework of the SFB TR6 / project C4 of the Deutsche Forschungsgemeinschaft (DFG). It was partially supported by the MAINZ graduate school. We are grateful to I. Snook for helpful and stimulating discussions.

\end{document}